\documentclass[intlimits,twoside,a4paper]{article}

\usepackage[cp1251]{inputenc}

\usepackage[eqsecnum]{cmpj3}

\issue{2024}{27}{1}{14701}
\doinumber{10.5488/CMP.27.14701}

\title[Description of CuInP$_{2}$S$_{6}$ ferrielectrics in a mixed Ising model]%
{Description of CuInP$_{2}$S$_{6}$ ferrielectrics in a mixed Ising model}
\author[R. Yevych, V. Liubachko, V. Hryts, M. Medulych, A.~Kohutych, Yu. Vysochanskii]{R. Yevych\orcid{0000-0001-6971-3429}, V. Liubachko\orcid{0000-0002-5114-2821} \thanks{Corresponding author: \email{vitalii.liubachko@uzhnu.edu.ua}.},\,\,\,V. Hryts\orcid{0009-0007-0836-7367}, M. Medulych\orcid{0000-0003-2128-0235}, A.~Kohutych\orcid{0000-0001-9138-2267}, \mbox{Yu. Vysochanskii}\orcid{0000-0002-2501-1780}}
\address{Institute for Solid State Physics and Chemistry, Uzhhorod National University, 3 Narodna Sq., 88000 Uzhhorod, Ukraine}


\date{Received February 13, 2024, in final form February 29, 2024}

\begin{document}

\maketitle

\begin{abstract}
The appearance of spontaneous polarization in CuInP$_{2}$S$_{6}$ ferrielectrics is related to the second order Jahn-Teller effect for copper cations located in a double-well local potential, the stereoactivity of indium cations located in a three-well local potential, as well as the valence fluctuations of phosphorus cations. The paraelectric to ferrielectric phase transition is primarily determined by the coupling of indium cations with their surroundings. This transition can be analyzed using the mixed Ising model with spins $s = 1/2$ and $S = 1$. The spectrum of pseudospin fluctuations at different temperatures was calculated using a mean-field approach for a set of quantum anharmonic oscillators. The results were then compared with Raman spectroscopy data for CuInP$_{2}$S$_{6}$ crystal. The analysis indicates that the lattice anharmonicity below 150 K, is mainly determined by the indium sublattice, leading to the coexistence of the glassy state and ferrielectric phase. Above 150 K, the anharmonicity of the copper sublattice activates the ionic conductivity and results in the existence of a long-ranged fluctuated cluster of spontaneous polarization in a temperature interval of the paraelectric phase above $T_{{C}}$.

\printkeywords
\end{abstract}

The temperature behavior of the Sn$_{2}$P$_{2}$S$_{6}$ ferroelectric was explained within the Blume-Capel model with pseudospin $S = 1$ related to the local three-well potential for the spontaneous polarization fluctuations~\cite{Yevych2016,Velychko2019,Erdem2022}. In the case of the CuInP$_{2}$S$_{6}$ crystal, the mixed Ising model \cite{Selke2010,Selke2011} with spins \mbox{$s = 1/2$} and $S = 1$, that are related to the ordering dynamics of Cu$^{+}$ cations in the double-well potential and to the ordering/displacive dynamics of In$^{3+}$ cations in the three-well local potential can be applied.

A temperature-pressure phase diagram within the quantum anharmonic oscillator (QAO) model has already been used to describe the Sn$_{2}$P$_{2}$S$_{6}$ ferroelectric crystal \cite{Yevych2018}. The results of this modelling are consistent with the consideration of the presence of the tricritical point on the temperature-pressure or temperature-composition diagrams within the Blume-Capel model \cite{Yevych2016}. For CuInP$_{2}$S$_{6}$ ferrielectrics, the thermodynamic properties have also been simulated within the QAO model \cite{Vysochanskii2023, Yevych2024}. In this Rapid Communication, we elucidate the different roles of Cu$^{+}$ ordering and In$^{3+}$ ordering/displacive dynamics in the spontaneous polarization of the CuInP$_{2}$S$_{6}$ ferrielectric crystals by studying the temperature dependence of the pseudospin fluctuation spectra within the QAO model. The results of these calculations are compared with the data from the light scattering study and are combined with the discussion of the observed anharmonic anomalies for the CuInP$_{2}$S$_{6}$ crystal lattice with the predictions of the mixed Ising model with spins $s = 1/2$ and \mbox{$S = 1$} . The analysis of the temperature-pressure phase diagram of CuInP$_{2}$S$_{6}$ ferrielectrics will be published further.

\begin{figure}[!b]
	\centerline{\includegraphics[width=0.8\textwidth]{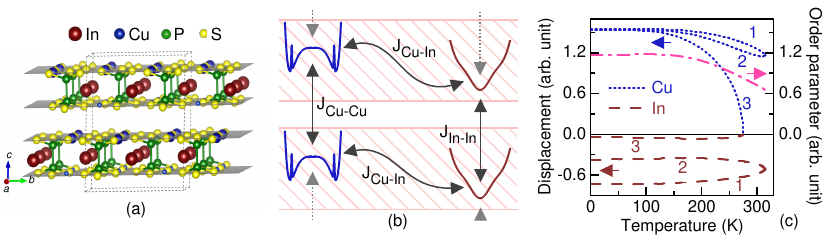}}
	\caption{(Colour online) (a) Crystal structure, (b) the QAO model and (c) the temperature dependence of the order parameter (spontaneous polarization) in the ferrielectric phase (pink dashed dotted line) as the sum of the oppositely oriented polarization contributed by Cu$^{+}$ cations (blue dotted line) and In$^{3+}$ cations (brown dashed line) for CuInP$_{2}$S$_{6}$ crystal. Solutions \textbf{1} and \textbf{2} are related to the first order phase transition at \mbox{$T_{C}$ = 315 K}. The metastable solution \textbf{3} is related to the second order phase transition at \mbox{$T_{\textrm{0}}$ = 275 K}.} \label{FIGURE1}
\end{figure}

During cooling from the paraelectric phase, the CuInP$_{2}$S$_{6}$ crystal [figure \ref{FIGURE1} (a)] undergoes a first order ferrielectric phase transition at about 315 K \cite{Maisonneuve1997}. This transition results in a symmetry decrease from $\mathit{C}$2$\mathit{c}$ to \textit{Cc} due to the ordering of copper cations in a polar sublattice when the indium cations are slightly displaced in the opposite direction, forming a second polar sublattice \cite{Maisonneuve1997}. The primitive cell of the CuInP$_{2}$S$_{6}$ crystal consists of two formula units, and it can be represented in the QAO model \cite{Vysochanskii2023,Yevych2024} as two interacting sublattices of oscillators, with chains of equivalent centers for Cu$^{+}$ with a double-well local potential and In$^{3+}$ with a three-well local potential in two parallel structural layers [figure \ref{FIGURE1} (b)] with a Hamiltonian: $\mathcal{H}= \sum_{i}\left ( T\left ( x_{i} \right ) +V\left ( x_{i} \right )+J\left \langle x \right \rangle x_{i}\right)$, where $T\left ( x_{i} \right )$ and $V\left ( x_{i} \right )$ are operators of kinetic and potential energy, respectively, $\left \langle x \right \rangle$=$\sum_{n}p_{n}x_{n}$ represents the average value of the displacement coordinate $x_{i}$ and is proportional to the order parameter in the non-symmetric phase while in the paraelectric phase it equals zero, and $J$ is a coupling constant. Our system comprises two subsystems: one for Cu ions and another for In ions. Each subsystem includes the coupling constants $J_{\textrm{Cu-Cu}}$ and $J_{\textrm{In-In}}$, respectively [figure \mbox{\ref{FIGURE1} (b)}]. Additionally, there is a coupling constant $J_{\textrm{Cu-In}}$ between the two subsystems. The following coupling constant ratios were used in the calculations:  $J_{\textrm{In-In}}$/$J_{\textrm{Cu-Cu}}$ = $J_{\textrm{In-In}}$/$J_{\textrm{Cu-In}}$ $\approx$ 12. The oscillator masses correspond to the atomic masses of copper and indium. The mean-field approach is reflected by the last term in the Hamiltonian with taking the relation $\sum_{ij}J_{ij}x_{i}x_{j}\approx \sum_{i}J\left \langle x \right \rangle x_{_{i}}$. 

In the QAO model, the symmetry-breaking field acting on the effective particle is calculated self-consistently. The Schr\"odinger equation is solved using total Hamiltonian to obtain a set of eigenenergies and their wave functions. The last were then used to calculate the average expectation value. The total Hamiltonian includes a temperature based on the Boltzmann distribution. The local potential for Cu$^{+}$ cations was modelled using an even power function of the displacement $x$: $U = -ax$$^{8}$+$bx$$^{10}$. The coefficients were chosen to accurately represent the temperature disordering of copper cations as observed by electron diffraction \cite{Maisonneuve1997}. To model the potential barrier associated with the thermal throw of copper cations into the interlayer van der Waals space, a Gaussian-like term was added to the copper local potential. The addition of this term to the potential, which in fact turned it into a four-well potential, only affected the calculated spatial distribution of copper cations (the appearance of the mentioned barrier) and had no significant effect on the other results presented in this work. To simulate the ordering/displacive dynamics of In$^{3+}$ cations, a three-well local potential with metastable side wells was used. Based on the structural data \cite{Maisonneuve1997}, the displacements from the center of the structural layer at \mbox{0 K} were normalized by a deviation of \mbox{1.55 $\textup{~\AA}$} for Cu$^{+}$ cations and 0.24 $\textup{~\AA}$ for In$^{3+}$ cations. The calculated displacements were then multiplied by the nominal ionic charges to estimate the contributions of the cations to the spontaneous polarization. More specifically, the application of the QAO model to this type of crystals has already been considered in references \cite{Yevych2016,Yevych2018,Yevych2024}.

\begin{figure}[!t]
	\centerline{\includegraphics[width=0.9\textwidth]{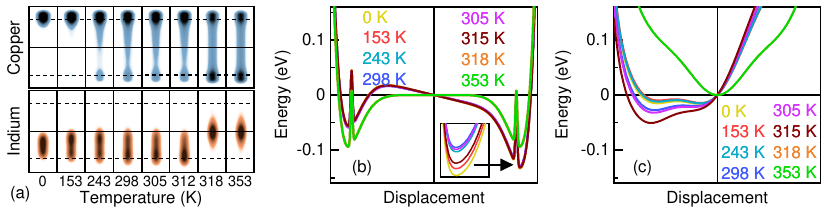}}
	\caption{(Colour online) Probability density contours for the spatial distribution of copper and indium cations for solution \textbf{2} in the QAO model (a). Temperature transformation across the ferrielectric to paraelectric phase transition of the double-well potential for Cu$^{+}$ cations (b) and the three-well potential for In$^{3+}$ cations (c).} \label{FIGURE2}
\end{figure}

\begin{figure}[!t]
	\centerline{\includegraphics[width=0.9\textwidth]{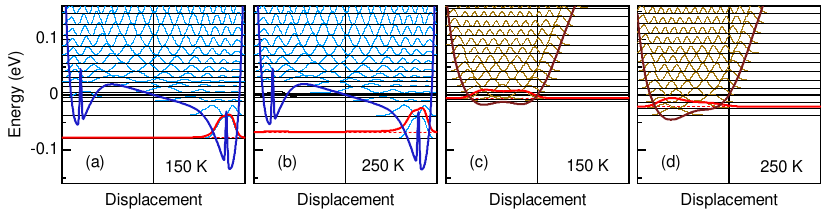}}
	\caption{(Colour online) Double-well local potential for Cu$^{+}$ cations (a, b) and three-well local potential for In$^{3+}$ cations (c, d) in CuInP$_{2}$S$_{6}$ crystal at 150 K and 250 K (thick line) and their energy levels with the corresponding square wave functions (thin lines). According to the probabilities $P_{{n}}$ of finding the oscillator at the $n$-th level with the energy $E_{{n}}$ at a given temperature, the average energy levels and wave functions are also calculated (red lines). } \label{FIGURE3}
\end{figure}

The temperature transformations of the spatial distribution of Cu$^+$ cations (related to the first order ferrielectric to paraelectric phase transition) for two solutions:  \textbf{1} and  \textbf{2} [figure \ref{FIGURE1} (c)], which coincide at \mbox{$T_{{C}}$ = 315 K}, are similar. With the selected parameters of the model they agree
%
%
with the experimentally observed \cite{Maisonneuve1997} polar distribution at low temperatures and symmetrical distribution at top and bottom positions within the structural layers above $T_{{C}}$ (figure \ref{FIGURE2}). The In$^{3+}$ cations at low temperatures in the case of the solution \textbf{1} are displaced out of the center of the structural layers, but for the slightly more energetic solution \textbf{2}, their dynamics is determined by the potential with two wells that are separated by a small barrier. The spatial distribution of In$^{3+}$ cations is very smeared and shows complex transformations with heating to $T_{{C}}$, as determined by the changes in its local potential [figure \ref{FIGURE2} (c)].

The temperature transformation of the spectra corresponding to the transitions between the energy
	levels of the quantum oscillator with a three-well potential, associated with In$^{3+}$ cations, and of
	the quantum oscillator with a double-well potential, associated with Cu$^{2+}$ cations, is shown in
 figures~\ref{FIGURE3} and~\ref{FIGURE4}.
At low temperatures, for the Cu$^{+}$ quantum oscillator, the transitions between the levels are excited with an energy difference of about 350 cm$^{-1}$ which is evident from the appearance of isolated spectral lines. With heating above 150 K and approaching the first order phase transition ($T_{{C}}$ = 315 K), for both solutions \textbf{1} and \textbf{2}, the additional spectral line appears near 230 cm$^{-1}$, and further, above 250 K, the spectral weight passes to a group of overlapping lines in the frequency range of 50$-$170 cm$^{-1}$. There is a rich set of spectral lines with the frequencies above 30 cm$^{-1}$ in the paraelectric phase, which do not show any significant shift with increasing temperature.

The spectrum corresponding to the transitions between the energy levels of the quantum oscillator,
associated with In$^{3+}$ cations, shows differences for solutions 1 and 2 (see figure \ref{FIGURE4} and \ref{FIGURE5}).
For the solution \textbf{1}, a clear \textit{displacive} character is illustrated by the frequency shift of the spectral lines from 180 cm$^{-1}$ in the ground state to 80 cm$^{-1}$ near $T_{{C}}$. Here, the additional lower frequency spectral lines already appear at 80 K and a pronounced frequency softening starts above 200~K. 

For the solution \textbf{2}, at 0 K, the spectral line associated with the transitions between the first and the second energy levels of the In$^{3+}$ quantum oscillator is located at the lowest frequency near 40 cm$^{-1}$. With a slight heating, above 20 K, an additional higher frequency spectral line appears, and at 150 K, the spectral weight is shifted to the group of lines concentrated near 100 cm$^{-1}$. Such a temperature evolution shows \textit{order-disorder} characteristics of the dynamics of the In$^{3+}$ cations at very low temperatures below 150~K.

\begin{figure}[!b]
	\centerline{\includegraphics[width=0.9\textwidth]{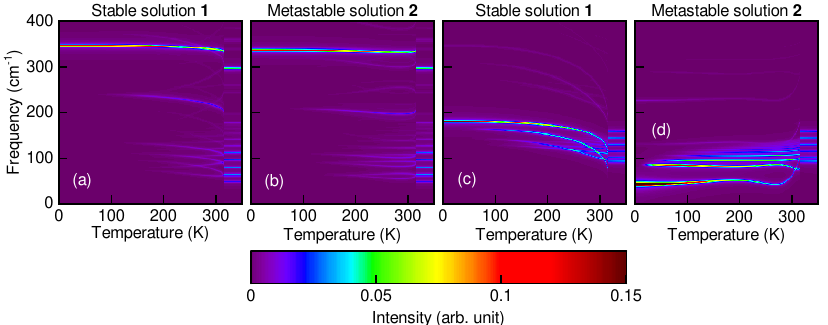}}
	\caption{(Colour online) Temperature dependence of the pseudospin fluctuation spectra calculated in the QAO model for Cu$^{+}$ cations (a, b) and for In$^{3+}$ cations (c, d) for CuInP$_{2}$S$_{6}$ crystal.} 
	\label{FIGURE4}
\end{figure}

\begin{figure}[!t]
	\centerline{\includegraphics[width=0.9\textwidth]{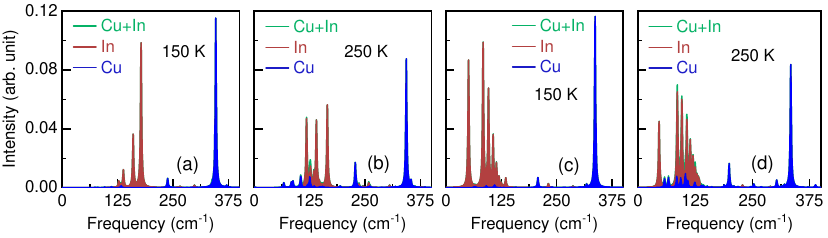}}
	\caption{(Colour online) Pseudospin fluctuation spectra calculated in the QAO model for Cu$^{+}$ cations (blue) and for In$^{3+}$ cations (brown) in the case of the stable solution \textbf{1} (a, b) and metastable solution \textbf{2} (c, d) at 150 K and 250 K for CuInP$_{2}$S$_{6}$ crystal.} \label{FIGURE5}
\end{figure}

The temperature evolution of the energy spectrum for a system of anharmonic quantum oscillators generally agrees with the tendency of the temperature changes in experimentally observed Raman spectra in CuInP$_{2}$S$_{6}$ crystal. At heating in the ferrielectric phase, low energy spectral lines slightly decrease their frequency and significantly change their width and shape asymmetry [figure \ref{FIGURE6} (a)]. The intensity of the spectral lines below 35 cm$^{-1}$ is very small at low temperatures. These spectral lines show some softening in frequency with heating to 150 K. Above 150 K, the observed Raman spectrum becomes more smeared, with widened contours of separated lines. The observed temperature evolution of the CuInP$_{2}$S$_{6}$ low frequency Raman spectra below 150 K 
can be related to the calculated within the QAO model (\ref{FIGURE4} and \ref{FIGURE5}) mixed
 disordering/displacive dynamics of the In$^{3+}$ cations over the low temperature range of the ferrielectric phase. At higher temperatures, the thermally activated relaxation dynamics of the Cu$^{+}$ cations is thermally activated, and the developed structural disordering induces the smearing of the Raman spectral lines.

\begin{figure}[htb]
	\centerline{\includegraphics[width=0.9\textwidth]{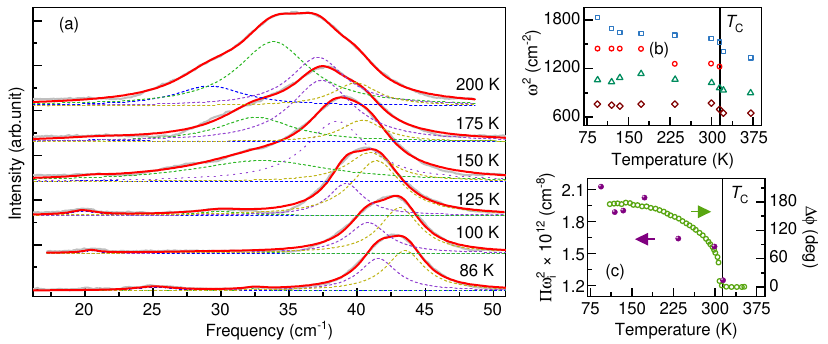}}
	\caption{(Colour online) (a) Low frequency Raman spectra of CuInP$_{2}$S$_{6}$ crystal at different temperatures. Dashed contours show separated spectral lines. (b) Temperature dependence of the squared frequencies of the Raman spectral lines across the ferrielectric phase transition in CuInP$_{2}$S$_{6}$ crystal. (c) Temperature dependence of the product of the squared frequencies for four low energy spectral lines (purple spheres). Green empty circles show the temperature dependence of the rotation of the light polarization \cite{Vysochanskii2003}, which is proportional to the square of the spontaneous polarization in the ferrielectric phase of CuInP$_{2}$S$_{6}$ crystal.} \label{FIGURE6}
\end{figure}

\begin{figure}[!t]
	\centerline{\includegraphics[width=0.95\textwidth]{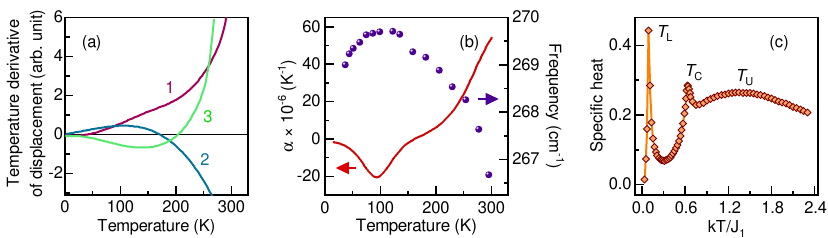}}
	\caption{(Colour online) (a) Change with heating of the temperature derivatives of the displacements of In$^{3+}$ cations for the solutions \textbf{1}, \textbf{2} and \textbf{3} according to the calculations in the QAO model. (b) Temperature dependence of the linear thermal expansion coefficient $\alpha$ \cite{Song2023} together with the temperature evolution of the lattice vibration frequency for CuInP$_{2}$S$_{6}$ crystal \cite{Song2023}. (c) Temperature dependence of the specific heat $\textit{C}$($\textit{T}$) calculated within a mixed Ising model with two non-critical anomalies at temperatures $T_{\textrm{L}}$ and $T_{\textrm{U}}$, and a critical anomaly at $T_{{C}}$ related to the phase transition \cite{Selke2011}.} \label{FIGURE7}
\end{figure}

According to the temperature evolution of the Raman spectra [figure \ref{FIGURE6} (a)], the product of the squared frequencies for several lowest energy optical modes with heating in the ferrielectric phase and approaching the first order phase transition temperature of the CuInP$_{2}$S$_{6}$ crystal correlates with the square of the order parameter (spontaneous polarization), which is proportional to the optical birefringence temperature behavior \cite{Vysochanskii2003} [figure \ref{FIGURE6} (b, c)]. 

The evolution of the temperature derivatives of the displacement of indium cations upon heating [figure \ref{FIGURE7} (a)] correlates with the qualitative change of the CuInP$_{2}$S$_{6}$ structural evolution at cooling below 150 K (change from a positive to a negative value of the thermal expansion coefficient \cite{Song2023}, change of the temperature behavior of the lattice vibrations frequency \cite{Song2023} [figure \ref{FIGURE7} (b)] and the coexistence of the ferrielectric phase with a dipole glass state at low temperatures \cite{Macutkevic2008}). This demonstrates the importance of the disordering/displacive dynamics of the In$^{3+}$ cations for the physical properties of CuInP$_{2}$S$_{6}$ ferrielectrics.

Using Monte Carlo simulations within the mixed Ising model with spins $s = 1/2$ and $S = 1$, the temperature dependence of the specific heat \textit{C}(\textit{T}) shows a three-peak structure, with two non-critical maxima and the critical peak in between [figure \ref{FIGURE7} (c)] \cite{Selke2010,Selke2011}. The sharp, but non-critical anomaly at low temperatures ($T_{\textrm{L}}$) results from the flipping $S = 1$ spins to the 0 state, while the broad non-critical maximum at high temperatures ($T_{\textrm{U}}$) results from the thermal activation of sufficiently large clusters of \mbox{$s = 1/2$} spins that persist above the phase transition. Between these temperatures, the critical peak associated with the transition from the ordered polar phase to the disordered one is presented \cite{Selke2011}. Projecting these simulation results onto the CuInP$_{2}$S$_{6}$ ferrielectric system, the non-critical peak at $T_{\textrm{L}}$ shows the partial disordering of the In$^{3+}$ sublattice, with S spins thermally flipped from the $+1$ or $-1$ state to 0. Between the two non-critical maxima in the $\textit{C}$($\textit{T}$) dependence, a critical peak at $T_{{C}}$ shows the transition where both sublattice polarizations vanish. The origin of the maximum at $T_{\textrm{U}}$ is due to the fact that at the critical point, the \mbox{$s = 1/2$} spins on the Cu$^{+}$ sublattice form rather large clusters of different orientations, leading to the vanishing of the order parameter. These clusters shrink rapidly near $T_{\textrm{U}}$, due to thermally activated flipping of the \mbox{$s = 1/2$} spins. 

Observed by Brillouin spectroscopy \cite{Kohutych2022} and according to ultrasonic data \cite{Samulionis2009}, the elastic softening of the longitudinal acoustic waves for the CuInP$_{2}$S$_{6}$ crystals at the edge of stability of the paraelectric phase can be induced by the inhomogeneously polarized state, which could have the same origin as the noncritical maximum of $\textit{C}$($\textit{T}$) at $T_{\textrm{U}}$. The coexistence of the dipole glassy state with the ferrielectric phase in CuInP$_{2}$S$_{6}$ crystals at low temperatures \cite{Macutkevic2008,Vysochanskii2003} can be interpreted as the evidence of disordering related to the non-critical maximum at $T_{\textrm{L}}$ on the $\textit{C}$($\textit{T}$) dependence. These experimental data confirm the importance of further research on the CuInP$_{2}$S$_{6}$ type ferrielectrics within the mixed Ising model.

\bibliographystyle{cmpj}
\bibliography{cmpjxampl}

\ukrainianpart

\title{Опис сегнетиелектрика CuInP$_{2}$S$_{6}$ у змішаній моделі Ізинґа}
\author{Р. Євич, В. Любачко, В. Гриць, М. Медулич, А. Когутич, Ю. Височанський}
\address{Інститут фізики і хімії твердого тіла, Ужгородський національний університет, пл. Народна 3, 88000 Ужгород, Україна
}

\makeukrtitle

\begin{abstract}
\tolerance=3000%
Виникнення спонтанної поляризації в сегнетиелектрику CuInP$_{2}$S$_{6}$ пов'язане з вториним ефектом Яна-Теллера для катіонів міді, розташованих у двоямному локальному потенціалі, стереоактивністю катіонів індію, розташованих у триямному локальному потенціалі, а також з валентними флуктуаціями катіонів фосфору.  Фазовий перехід з параелектричної в сегнетиелектричну фазу визначається насамперед зв'язком катіонів індію з їхнім оточенням. Цей перехід можна проаналізувати за допомогою змішаної моделі Ізинґа зі спінами $s = 1/2$ і $S = 1$. Спектр флуктуацій псевдоспінів при різних температурах розраховано в рамках моделі квантових ангармонічних осциляторів у наближенні середнього поля. Результати порівняні з даними раманівської спектроскопії для кристала CuInP$_{2}$S$_{6}$. Аналіз показує, що ангармонічність ґратки нижче 150 К в основному визначається підґраткою індію, що зумовлює співіснування склоподібного стану і сегнетиелектричної фази. Вище 150 К ангармонічність підґратки міді активує іонну провідність і зумовлює існування великорозмірних динамічних кластерів спонтанної поляризації в температурному інтервалі параелектричної фази вище $T_{{C}}$.

\keywords сегнетиелектрики, CuInP$_{2}$S$_{6}$, змішана модель Ізинґа, модель квантових ангармонічних осциляторів

\end{abstract}

\end{document}